\documentclass[manuscript]{aastex}
\usepackage[numberedappendix]{emulateapj5}

\slugcomment{To appear in The Astrophysical Journal}

\newcommand\xx{{\bf x}}
\newcommand\uu{{\bf u}}
\newcommand\mat[4]{\left[\begin{array}{cc} #1 & #2 \cr #3 & #4 \cr \end{array}\right]}
\renewcommand\th{\theta}
\newcommand\Q{{\cal Q}}

\newcommand\refeqs[2]{eqs.~(\ref{eq:#1}) and (\ref{eq:#2})}
\newcommand\reffig[1]{Figure~\ref{fig:#1}}

\makeatletter
\newenvironment{figureone}
  {\def\@captype{figure}}
  {}

\makeatother

\begin{document}

\title{Source Ellipticity and the Statistics of Lensed Arcs}
\author{Charles R.\ Keeton}
\affil{Steward Observatory, University of Arizona \\
  933 N.\ Cherry Ave., Tucson, AZ 85721}
\affil{\it and}
\affil{Astronomy and Astrophysics Department, University of Chicago \\
  5640 S.\ Ellis Ave., Chicago, IL 60637}

\begin{abstract}
The statistics of gravitationally lensed arcs, which can be
used for a variety of cosmological tests, are sensitive to the
intrinsic shapes of the source galaxies.  I present an analytic
formalism that makes it simple to include elliptical sources
in analytic calculations of lens statistics.  For cuspy lens
models, sources with an axis ratio of 2:1 enhance the total
number of arcs longer than 10:1 by a factor of order two, while
modestly {\it decreasing\/} the number ratio of radial arcs to
tangential arcs.  Source ellipticity is therefore an important
systematic effect in detailed quantitative studies, but it
should not hinder cosmological applications such as attempts
to constrain cluster dark matter profiles with arc statistics.
\end{abstract}


\section{Introduction}

A cluster of galaxies can distort background galaxies into long,
thin arcs by gravitational lensing.  The statistics of lensed
arcs can be used to constrain cluster mass distributions (e.g.,
Wu \& Hammer 1993; Miralda-Escud\'e 1993a; Bartelmann, Steinmetz
\& Weiss 1995; Molikawa \& Hattori 2000), the population of faint,
distant galaxies (e.g., Miralda-Escud\'e 1993b; Hamana \& Futamase
1997; B\'ezecourt, Pell\'o \& Soucail 1998), and cosmological
parameters (e.g., Wu \& Mao 1996; Bartelmann et al.\ 1998).
Comparing the number of arcs extended radially relative to the
cluster center with the number of arcs stretched tangentially
may be an especially good probe of the central density profile
in clusters (Molikawa \& Hattori 2000; Oguri, Taruya \& Suto
2001).

Most calculations of arc statistics assume that the source
galaxies are circularly symmetric on the sky.  The exception
is a series of papers by Bartelmann et al.\ (1995, 1998; also
Meneghetti et al.\ 2000), which show that elliptical sources
increase the likelihood of long, thin arcs but do not carefully
quantify the effect.  As applications of arc statistics become
increasingly detailed, it is important to revisit the question
of how elliptical sources affect the statistics of both tangential
and radial arcs.  In this paper I develop an analytic formalism
for including elliptical sources in studies of arc statistics,
which makes elliptical sources as easy to work with as circular
sources.  Section 2 presents the formalism, Section 3 gives
examples to quantify the effects of elliptical sources, and
Section 4 offers conclusions.

\section{Analytic Formalism}

The lens equation relates the angular position $\xx$ of a lensed
image to the angular position $\uu$ of the intrinsic source,
\begin{equation}
  \uu = \xx - \nabla\phi(\xx)\,,
\end{equation}
where $\phi$ is the projected gravitational potential of the
lens, which is related to the projected surface mass density
$\Sigma(\xx)$ via the two-dimensional Poisson equation
$\nabla^2\phi(\xx) = \Sigma(\xx)/\Sigma_{cr}$, where $\Sigma_{cr}$
is the critical surface density for lensing (e.g., Schneider,
Ehlers \& Falco 1992).  The distortions of an image are
determined by the magnification tensor
\begin{equation}
  \mu = \mat{1-\phi_{xx}}{-\phi_{xy}}{-\phi_{xy}}{1-\phi_{yy}}^{-1} ,
\end{equation}
where $\phi_{ij} = \partial^2\phi/\partial x_i \partial x_j$ are
second derivatives of the potential.

Consider a source with elliptical symmetry, so it is described
by a surface brightness distribution with the form $S(\uu) =
f(\rho_s)$, where $\rho_s$ is an ellipse coordinate given by
\begin{eqnarray}
  \rho_s^2 &=& \uu^t \cdot R^t \cdot \Q_s \cdot R \cdot \uu\,,
    \label{eq:rhos} \\
  \mbox{where}\quad
  \Q_s &=& \mat{1\ }{0}{0\ }{q_s^2} , \\
  R &=& \mat{\cos\th\ }{\sin\th}{-\sin\th\ }{\cos\th} ,
\end{eqnarray}
where $q_s \ge 1$ is the axis ratio of the source ellipse, $\th$
is the orientation angle of the ellipse, $R$ is a rotation matrix,
and $t$ denotes the matrix transpose.  Without loss of generality,
we can choose coordinates aligned with the eigenvectors of the
lensing magnification tensor, so $\th$ is the angle between the
source ellipse and the dominant eigenvector of the magnification
tensor (the shear vector). In this coordinate system, the
magnification tensor can be written as
\begin{equation}
  \mu = a\,\mat{q_l}{0}{0}{1} ,
\end{equation}
where $q_l \ge 1$ is the ratio of the magnification eigenvalues;
it gives the axis ratio of an image from a circular source, so I
refer to it as the lensing axis ratio. Also, $a$ is a scale factor
that gives the uniform expansion or contraction of the image
relative to the source.

Analytic studies of arc statistics often assume a small source
so the shape of the image is easy to calculate, and the assumption
appears to be broadly valid (see Hattori, Watanabe \& Yamashita 1997).
In this limit, the image is an ellipse described by the surface
brightness distribution $I(\xx) = S(\mu_{inv}\cdot\xx) = f(\rho_i)$,
where $\mu_{inv}$ is the inverse of the magnification tensor, and
$\rho_i$ is an ellipse coordinate given by
\begin{eqnarray}
  \rho_i^2 &=& \xx^t \cdot M \cdot \xx\,, \\
  \mbox{where}\quad
  M &=& \mu_{inv}^t \cdot R^t \cdot \Q_s \cdot R \cdot \mu_{inv}\,.
\end{eqnarray}
(Compare eq.~\ref{eq:rhos}.)
The axis ratio of the observed image, $q_{obs} \ge 1$, is
given by the eigenvalues of the tensor $M$ to be
\begin{eqnarray}
  q_{obs} &=& \left[ { {T + \left(T^2-4D\right)^{1/2}} \over
    {T - \left(T^2-4D\right)^{1/2}} } \right]^{1/2} , \\
  \mbox{where}\quad
  T &=& q_l^2 + q_s^2 + (q_l^2-1)(q_s^2-1)\cos^2\th\,, \\
  D &=& q_l^2\,q_s^2\,.
\end{eqnarray}
Some simple limits are as follows:
\begin{equation}
  q_{obs} = \cases{
    q_l & if $q_s=1$ \cr
    q_l\,q_s & if $\th=0$ \cr
    \max\left({q_l \over q_s},{q_s \over q_l}\right) & if $\th=\pi/2$ \cr
  }
\end{equation}

When computing the cross section for arcs, each source position
must be weighted by the probability that a source at that position
produces a lensed arc.  An arc is defined as an image with an
axis ratio larger than some threshold, i.e., $q_{obs} \ge Q$
for some value of $Q$; a popular choice is $Q=10$ (e.g., Wu \&
Hammer 1993), although $Q=4$ may be more relevant for radial
arcs (see Oguri et al.\ 2001).  Consider a source with axis
ratio $q_s$ and orientation angle $\th$, at a position such that
the lensing axis ratio is $q_l$.  The source produces an arc if
\begin{eqnarray}
  \cos^2\th &\ge& \xi(Q,q_l,q_s)\,, \\
  \mbox{where}\quad
  \xi(Q,q_l,q_s) &=& { (Q^2+1)\,q_l\,q_s - Q(q_l^2+q_s^2)
    \over Q(q_l^2-1)(q_s^2-1) }\ . \label{eq:xi}
\end{eqnarray}
If the sources have random orientations, the probability of
obtaining an arc from a source at that position is then
\begin{equation}
  p_{arc}(Q,q_l,q_s) = \cases{
    (2/\pi)\,\cos^{-1}\sqrt{\xi(Q,q_l,q_s)} & if $0 \le \xi \le 1$ \cr
    1 & if $\xi<0$ \cr
    0 & if $\xi>1$ \cr
  } \label{eq:fsrc}
\end{equation}
\reffig{qplane} shows how to interpret this result. There are three
simple regions in the $(q_l,q_s)$ plane. First, for $q_l \ge
Q\,q_s$, the lensing distortion is strong enough that any source
at that position yields an arc ($p_{arc}=1$), regardless of the
source orientation. Second, for $q_s \ge Q\,q_l$, the sources are
intrinsically long enough that any source at that position produces
an arc (again $p_{arc}=1$), although this regime is uninteresting
because such elongated sources are rare, and because this case is
not really lensing. Third, for $q_l\,q_s < Q$ no source at that
position can produce an arc ($p_{arc}=0$), because neither the
intrinsic source shape nor the lensing distortion is large enough
(for any orientation).

\begin{figureone}
\centering\leavevmode\includegraphics[width=3.4in]{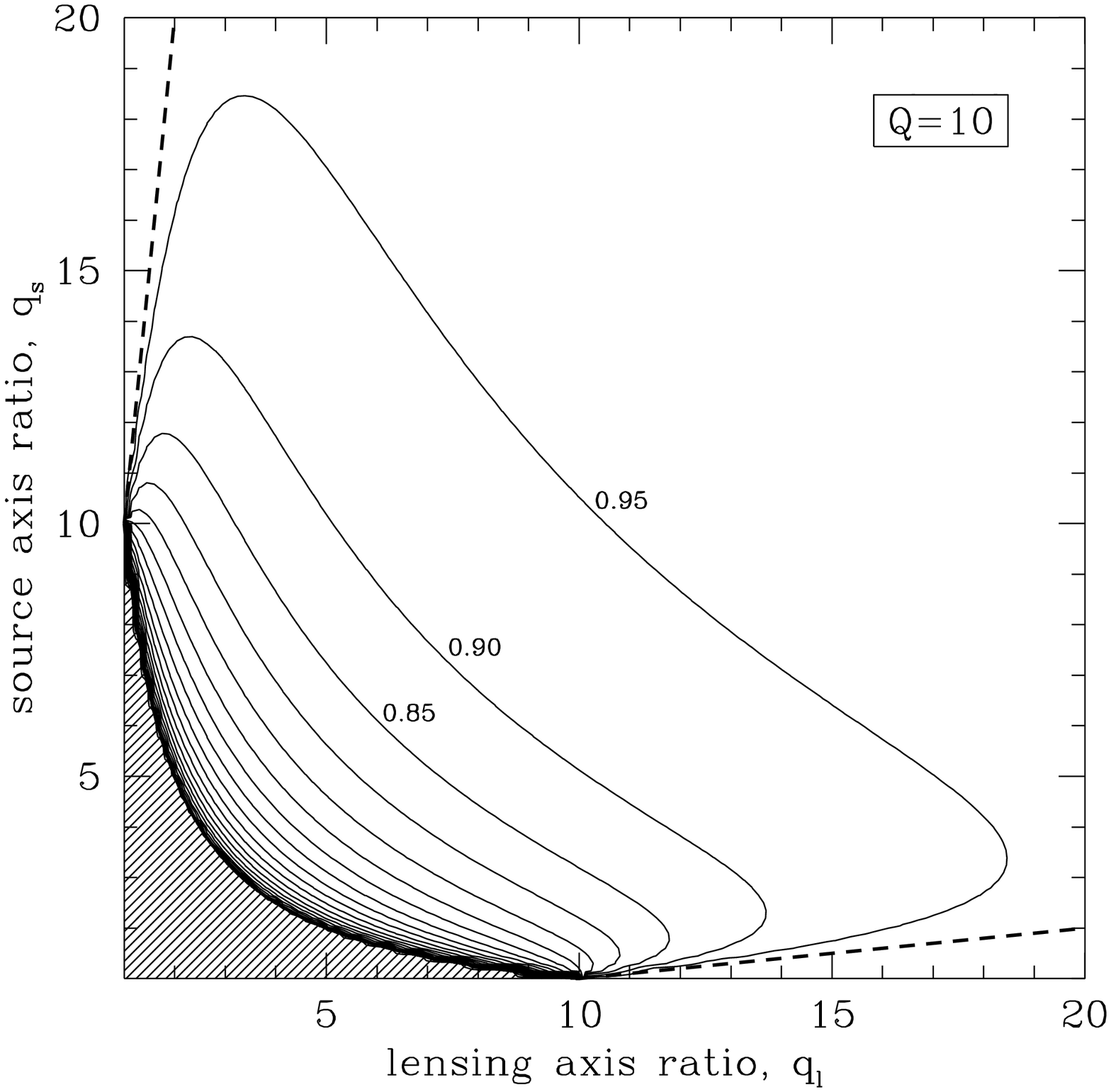}
\caption{
The arc probability, $p_{arc}$, as a function of $q_l$ and $q_s$,
for an arc threshold $Q=10$.  Contours are drawn at $p_{arc}=0.95,
0.90, 0.85,\ldots$ starting from the upper right, as indicated.
The heavy dashed lines indicate $p_{arc}=1$, while the shaded
region in the lower left indicates $p_{arc}=0$.  Source axis
ratios larger than $q_s \gtrsim 2$, while implausible, are included
to show that $p_{arc}$ is symmetric in $q_l$ and $q_s$.
}\label{fig:qplane}
\end{figureone}
\vspace{0.2in}

The remainder of the $(q_l,q_s)$ plane has intermediate values of
$p_{arc}$, with two interesting qualitative results. First, with
elliptical sources there are positions where the lensing distortion
is moderate ($q_l < Q$), but some sources can still produce arcs;
the arcs are produced when the sources are roughly aligned with
the lensing distortion. This result reproduces the result from
Bartelmann et al.\ (1995) that elliptical sources increase the
likelihood of long/thin arcs. Second, there are positions where
the lensing distortion is strong ($q_l > Q$), but not all sources
yield arcs; sources that are approximately orthogonal to the
lensing distortion fail to produce arcs.

This analysis makes it simple to include the effects of elliptical
sources in calculations of arc statistics. Given sources with a
fixed axis ratio $q_s$, the cross section for arcs with threshold
$Q$ is
\begin{equation}
  \sigma(Q,q_s) = \int d\uu\ p_{arc}\bigl(Q,q_l(\uu),q_s)\,,
  \label{eq:sig1}
\end{equation}
where $q_l(\uu)$ is the lensing distortion associated with source
position $\uu$, and the integral extends over the source plane.
While circular sources yield $p_{arc}=0$ or 1, elliptical sources
introduce variable weighting into the integral.  Because
\refeqs{xi}{fsrc} make the effect easy to compute, elliptical
sources can be added  to calculations of lens statistics with no
more work than circular sources. For source positions with multiple
images and hence multiple values of $p_{arc}$ (one for each image),
the largest value of $p_{arc}$ that corresponds to the desired type
of arc (radial or tangential) should be used.  The formalism can be
extended to a distribution of source shapes. Given a probability
distribution $P(q_s)$ for source shapes, the arc cross section is
\begin{equation}
  \sigma(Q) = \int d\uu \int dq_s\ p_{arc}\bigl(Q,q_l(\uu),q_s)\,P(q_s)\,.
  \label{eq:sig2}
\end{equation}
For each source position $\uu$, the integral over $q_s$ can be
computed with fast numerical integration.

\section{Examples}

In proposing that arc statistics can be used to constrain the
inner profiles of dark matter halos, Molikawa \& Hattori (2000)
and Oguri et al.\ (2001) consider only circular sources.  I
reexamine their proposal to give an example of the quantitative
effects of source ellipticity.  Following both studies, consider
a dark matter halo with a generalized NFW profile,
\begin{equation}
  \rho(r) = { \rho_0 r_0^3 \over r^{\gamma} (r_0+r)^{3-\gamma} }\ ,
\end{equation}
where $r_0$ is a scale radius and $\rho_0$ is a characteristic
density. The density profile has a central cusp $\rho \propto
r^{-\gamma}$; when $\gamma=1$ it corresponds to the NFW profile
(Navarro, Frenk \& White 1996, 1997), and when $\gamma=1.5$ it
resembles the steeper cusp advocated by, e.g., Fukushige \&
Makino (1997), Moore et al.\ (1998, 1999) and Klypin et al.\
(2000).  It is convenient to define a ``concentration'' parameter
$C = r_{200}/r_{*}$, where $r_{200}$ is the radius within which
the mean density is 200 times the critical density of the universe,
often taken to define the edge of halo (e.g., Crone, Evrard \&
Richstone 1994; Cole \& Lacey 1996; Navarro et al.\ 1996, 1997);
and $r_{*}$ is the radius at which the logarithmic slope of the
density is $d(\log\rho)/d(\log r)=-2$.  This definition is
equivalent to the standard concentration parameter for NFW halos,
and it is a good generalization to $\gamma \ne 1$ (see Keeton \&
Madau 2001).  Like Molikawa \& Hattori (2000) and Oguri et al.\
(2001), I consider spherical halos for simplicity, but the
formalism presented in Section 2 is fully general.

Lensing depends on the projected surface mass density in units
of the critical surface density for lensing.  For generalized NFW
halos, the lensing strength is determined by the dimensionless
parameter
\begin{eqnarray}
  \kappa_0 = { \rho_0 r_0 \over \Sigma_{cr} }
  &=& 10^{4/3}\ { (H^2 G M)^{1/3} \over c^2 }\ {D_{ol} D_{ls} \over
    D_{os} } \\
  &&\quad\times\ { (3-\gamma)(r_{200}/r_0)^{\gamma-1} \over {}_2
    F_{1}(3-\gamma,3-\gamma,4-\gamma,-r_{200}/r_0) }\ , \nonumber
\end{eqnarray}
where $M$ is the halo mass, $H$ is the Hubble parameter evaluated
at the lens redshift, and ${}_2 F_1(a,b,c,x)$ is the hypergeometric
function (e.g., Press et al.\ 1992). Also, $D_{ol}$, $D_{os}$, and
$D_{ls}$ are angular diameter distances from the observer to the
lens, from the observer to the source, and from the lens to the
source, respectively. The complete equations for generalized NFW
lenses are given by, e.g., Molikawa \& Hattori (2000) and Keeton
\& Madau (2001).

As a fiducial model, consider a lens of mass $10^{15}\, h^{-1}\,
M_\odot$ at redshift $z_l=0.3$, with sources at redshift $z_s=1$,
in a cosmology with matter density $\Omega_M=0.3$ and cosmological
constant $\Omega_\Lambda=0.7$. Such halos have a range of
concentration parameters with median $C \simeq 4.3$ (e.g., Eke,
Navarro \& Steinmetz 2000). The lensing strength parameter is
then:
\begin{equation}
  \kappa_0 = \cases{
    0.1776 & if $\gamma=1.0$ \cr
    0.0543 & if $\gamma=1.5$ \cr
  }
\end{equation}
Increasing or decreasing the strength parameter by $0.2$ dex
corresponds to changes in the halo mass of $0.6$ dex (a factor of
4), or to changes in the concentration of about $0.18$ dex (the
1$\sigma$ scatter in the concentration; Bullock et al.\ 2001).

\reffig{sig} shows how arc cross sections depend on the source
ellipticity $e_s=1-b/a$, where $a$ and $b$ are the semi-major
and semi-minor axes of the source, respectively. Increasing the
source ellipticity increases the cross section for both tangential
and radial arcs, but it affects tangential arcs more strongly.
From $e_s=0$ to $e_s=0.5$ the cross section for tangential arcs
increases by a factor of 2.6 for $\gamma=1.0$ models and 2.3 for
$\gamma=1.5$ models, while the cross section for radial arcs
increases by 1.7 for both models.  The enhancement factors are
not very sensitive to the lensing strength $\kappa_0$.  They
are somewhat more sensitive to assumptions about magnification
bias (which is not included in \reffig{sig}), but factors around
2 are still typical.

The statistic advocated by Molikawa \& Hattori (2000) is the
number ratio of radial and tangential arcs, which is shown in
\reffig{rat}.  Although both types of arcs are enhanced by
elliptical sources, radial arcs are affected less strongly;
hence, the ratio of radial to tangential arcs systematically
{\it decreases\/} with source ellipticity.  From $e_s=0$ to
$e_s=0.5$ the ratio decreases by a factor of about 1.5 for
$\gamma=1.0$ models and 1.4 for $\gamma=1.5$ models.  The
dependence on the strength parameter $\kappa_0$ is weak (which
is why the number ratio is an attractive statistic in the first
place).  Magnification bias is an important systematic effect
in the dependence of the arc number ratio on the cusp slope
$\gamma$ (see Oguri et al.\ 2001), but it does not dramatically
change the dependence on source ellipticity.

\vspace{0.2in}
\begin{figureone}
\centering\leavevmode\includegraphics[width=3.4in]{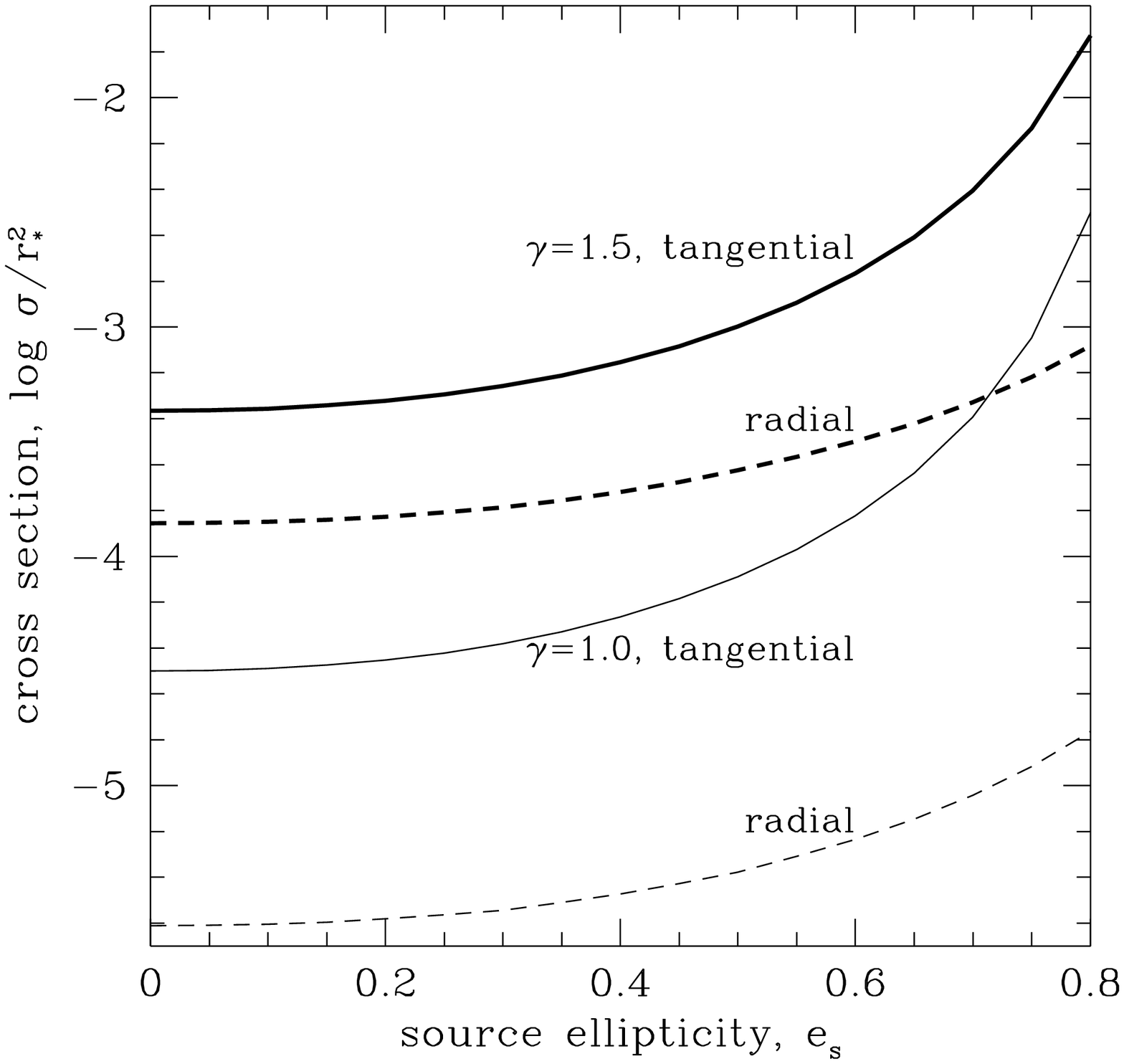}
\caption{
Arc cross sections versus source ellipticity, computed for the
fiducial model with a length/width threshold $Q=10$. The light
and heavy curves denote models with $\gamma=1.0$ and $\gamma=1.5$,
respectively. The solid and dashed curves indicate cross sections
for tangential and radial arcs, respectively. The cross sections
are computed in angular units and are normalized by the radius
$r_{*}$ at which the logarithmic slope of the density is $-2$.
}\label{fig:sig}
\end{figureone}

\begin{figureone}
\centering\leavevmode\includegraphics[width=3.4in]{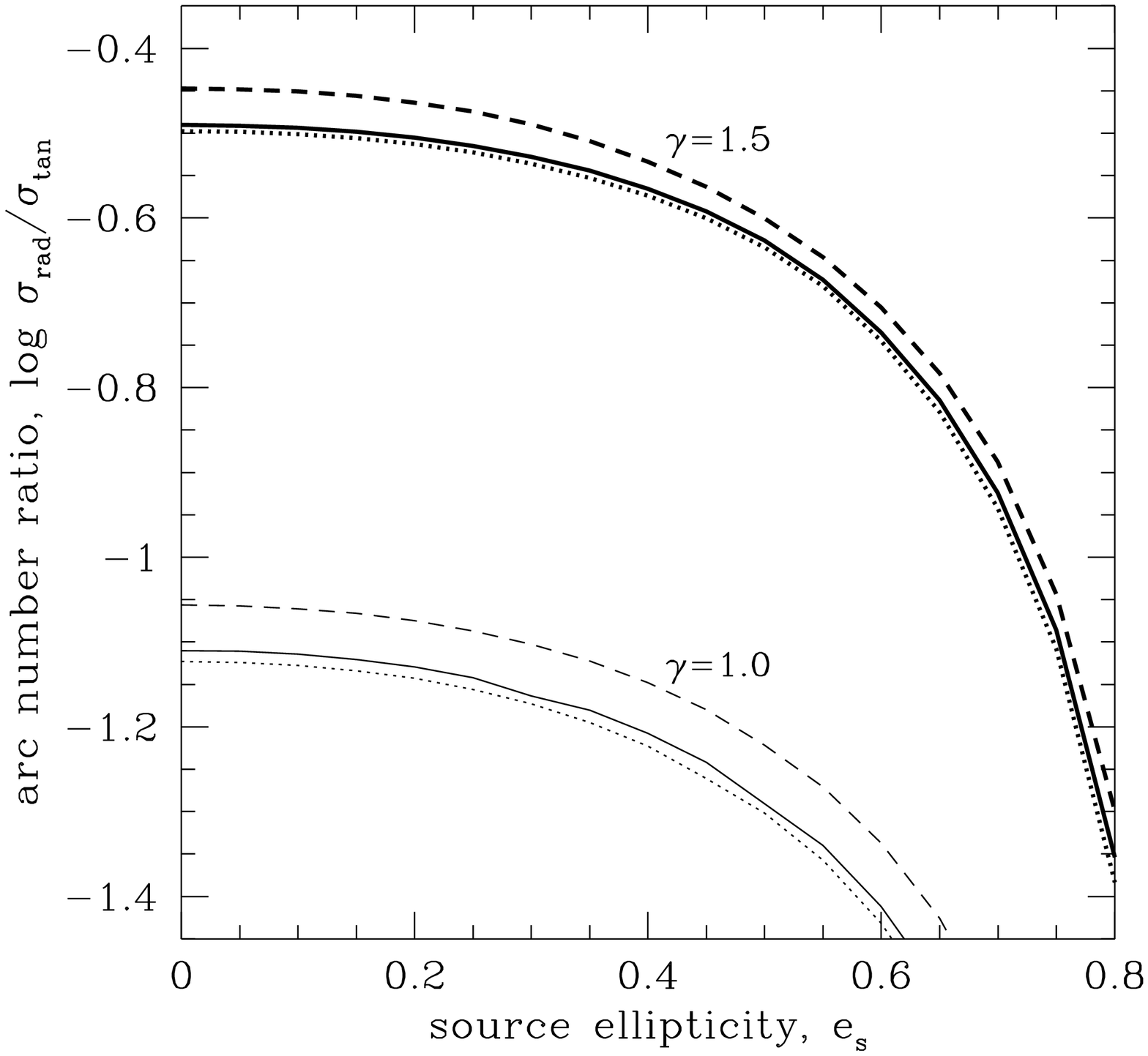}
\caption{
The number ratio of radial and tangential arcs as a function
of the source ellipticity, computed with a length/width threshold
$Q=10$.  The light and heavy curves denote models with $\gamma=1.0$
and $\gamma=1.5$, respectively.  The solid curves indicate the
fiducial model, while the dashed (dotted) curves indicate models
where the lensing strength is increased (decreased) by 0.2 dex.
}\label{fig:rat}
\end{figureone}
\vspace{0.2in}

For modest source ellipticities, $e_s \lesssim 0.5$, the change
in the arc number ratio due to source ellipticity is substantially
smaller than the difference between $\gamma=1.0$ and $\gamma=1.5$
models.  Hence, even if the source shape distribution is poorly
known, it should still be possible for arc statistics to
distinguish between models with different density cusps.
Source ellipticity is thus an important systematic effect that
must be included to obtain robust quantitative results, but it
is weaker than effects that one may wish to probe with arc
statistics.

These examples are based on axially symmetric lens models, as
were the studies by Molikawa \& Hattori (2000) and Oguri et al.\
(2001).  Departures from axial symmetry, such as an aspherical
cluster potential or subclumps in the cluster mass distribution,
can affect the number of arcs by an order of magnitude or more
(Bartelmann et al.\ 1995).  While asymmetry in the lens thus
appears to be more important than non-circular sources, the
quantitative effects of lens asymmetry on the arc number ratio,
and the combined effects of lens and source asymmetry, are
under investigation (M.\ Meneghetti et al., in preparation).

\section{Conclusions}

I have presented a simple formalism for including the effects
of elliptical sources in analytic calculations
of the statistics of lensed arcs.  It is straightforward to
compute, as a function of source position, the probability
that a source with a given axis ratio and a random orientation
yields a lensed arc.  This probability serves as a weighting
factor in the integral over source positions to determine the
lensing cross section.  The formalism is fully analytic, so
elliptical sources require no more work than circular sources;
and it is easily extended to a distribution of source shapes.

In generalized NFW lenses, elliptical sources enhance the
likelihood of producing both tangential and radial arcs;
sources with ellipticity $e_s=0.5$ increase the expected
number of arcs by about a factor of two (relative to circular
sources).  Radial and tangential arcs are affected differently,
so the number ratio of radial arcs to tangential arcs decreases
as $e_s$ increases; from $e_s=0$ to $e_s=0.5$ the ratio drops
by a factor of about 1.5.  This change is an important
systematic effect in quantitative studies of arc statistics,
but by itself it should not significantly limit attempts to
constrain the dark matter profile of clusters.  These
conclusions are drawn from examples with spherical lenses,
and it will be interesting to see whether elliptical sources
have any different effects when the lenses are non-spherical
(M.\ Meneghetti et al., in preparation).

\acknowledgements
Acknowledgements: I would like to thank Matthias Bartelmann for
interesting and helpful discussions.  This work was supported
by Steward Observatory.


\begin{references}

\reference{}
Bartelmann, M., Steinmetz, M., \& Weiss, A. 1995, \aap, 297, 1

\reference{}
Bartelmann, M., Huss, A., Colberg, J. M., Jenkins, A. \& Pearce, F. R.
1998, \aap, 330, 1

\reference{}
B\'ezecourt, J., Pell\'o, R., \& Soucail, G. 1998, \aap, 330, 399

\reference{}
Bullock, J. S., Kolatt, T. S., Sigad, T., Somerville, R. S.,
Kravtsov, A. V., Klypin, A. A., Primack, J. R., \& Dekel, A. 2001,
\mnras, 321, 559

\reference{}
Cole, S., \& Lacey, C. 1996, \mnras, 281, 716

\reference{}
Crone, M. M., Evrard, A. E., \& Richstone, D. O. 1994, \apj, 434, 402

\reference{}
Eke, V., Navarro, J. F., \& Steinmetz, M. 2000, preprint (astro-ph/0012337)

\reference{}
Fukushige, T., \& Makino, J. 1997, \apj, 477, L9

\reference{}
Hamana, T., \& Futamase, T. 1997, \mnras, 287, L7

\reference{}
Hattori, M., Watanabe, K., \& Yamashita, K. 1997, \aap, 319, 764

\reference{}
Keeton, C. R., \& Madau, P. 2001, \apj, 549, L25

\reference{}
Klypin, A., Kravtsov, A. V., Bullock, J. S., \& Primack, J. R. 2000,
preprint (astro-ph/0006343)

\reference{}
Meneghetti, M., Bolzonella, M., Bartelmann, M., Moscardini, L., \&
Tormen, G. 2000, \mnras, 314, 338

\reference{}
Miralda-Escud\'e, J. 1993a, \apj, 403, 497

\reference{}
Miralda-Escud\'e, J. 1993b, \apj, 403, 509

\reference{}
Molikawa, K., \& Hattori, M. 2000, preprint (astro-ph/0009343)

\reference{}
Moore, B., Governato, F., Quinn, T., Stadel, J., \& Lake, G. 1998,
\apj, 499, L5

\reference{}
Moore, B., Quinn, T., Governato, F., Stadel, J., \& Lake, G. 1999,
\mnras, 310, 1147

\reference{}
Navarro, J. F., Frenk, C. S., \& White, S. D. M. 1996, \apj, 462, 563

\reference{}
Navarro, J. F., Frenk, C. S., \& White, S. D. M. 1997, \apj, 490, 493

\reference{}
Oguri, M., Taruya, A., \& Suto, Y. 2001, preprint (astro-ph/0105248)

\reference{}
Press, W. H., Teukolsky, S. A., Vetterling, W. T., \& Flannery, B. P.
1992, Numerical Recipes in C: The Art of Scientific Computing,
Second Edition (New York: Cambridge Univ. Press)

\reference{}
Schneider, P., Ehlers, J., \& Falco, E. E. 1992, Gravitational Lenses
(New York: Springer)

\reference{}
Wu, X.-P., \& Hammer, F. 1993, \mnras, 262, 187

\reference{}
Wu, X.-P., \& Mao, S. 1996, \apj, 463, 404

\end{references}
\end{document}